\documentstyle[aps,prl,twocolumn]{revtex}

\begin{document}
\noindent{ {\bf Comment on ``Adsorption of Polyelectrolyte onto a Colloid of 
Opposite Charge''}}

In a recent Letter, Gurovitch and Sens \cite{GS} studied the adsorption of a weakly
charged polyelectrolyte chain onto an oppositely charged colloidal particle. 
By using a variational technique they found that the colloidal particle can adsorb a polymer
of higher charge than its own, and thus be ``overcharged.'' I argue that the observed
overcharging by a factor of $16/5$ is indeed an artifact of the approximations 
involved in the study. Moreover, I show that the existence of overcharging 
for a pointlike colloidal particle depends
crucially on the choice of the trial wave function, contrary to their claim.


To study the adsorption, they use a restricted class of trial wave functions 
$\psi_{z}({\bf r})$ based on the assumption that the polyelectrolyte
is {\it uniformly} confined in space to a sphere of size $1/z$, and treat $z$ as a
variational parameter. A finite value for $1/z$ that minimizes the 
free energy, called $1/z^*$,
would then mean complete adsorption,
whereas an infinite $1/z^*$ would imply instabilities in the form of dangling segments
stretching away from the core. 
I use a larger class of trial wave functions
\begin{equation}
\psi^2({\bf r})=\alpha \psi_{z}^2({\bf r})+{(1-\alpha) \over V},
\end{equation}
that assumes a fraction $\alpha$ of the chain is confined to a sphere of size
$1/z$, while the rest fills up a considerably larger space (of volume
$V$), and treat $z$ and $\alpha$ as parameters. This class of trial functions
clearly contains that used in Ref. \cite{GS} as a subclass ($\alpha=1$),
and can thus be used to check the robustness of their results. 

One can argue why the above choice for the trial wave function is
physically more appropriate. The polyelectrolyte can 
be either adsorbed to the oppositely charged colloid, or stretched out
due to self-repulsion. This suggests that an effective two dimensional
phase space is more suitable to describe the state of the system. Any configuration
of the chain can then effectively be described as a decomposition into
various segments, each of which occupying one of the two states,
in this simplified picture.
The natural question to ask is then the ``occupation'' ratio of each state,
which is determined by minimization of the free energy.

Consider a chain of length $N$ with a fraction $f$ of its monomers being
charged, which is adsorbed to a colloid of charge $-Q$.
Using $\psi_{z}({\bf r})=(z^3/\pi)^{1/2} e^{-zr}$ as in Ref. \cite{GS}
, one obtains the total free energy per unit charge as 
\begin{equation}
{E(z,\alpha) \over k_B T}=c_0 a^2 z^2 \alpha 
- c_1 Q l_b z \alpha + c_2 f N l_b z \alpha^2,
\end{equation}
in which $l_b=q^2/\epsilon k_B T$ is the Bjerrum length, $a^2=b^2/f$ where $b$
is the monomer size, and the numerical
coefficients are given as $c_0=1/6$, $c_1=1$, and $c_2=5/16$ for the above choice
of trial function. Minimizing with
respect to $z$ and $\alpha$ yields a ``confinement radius''
$1/z^*=3 c_0 a^2/c_1 l_b f Q$, and
a ``charging fraction''
\begin{equation}
\alpha^*=\left({c_1 \over 3 c_2}\right) \times {Q \over f N}.
\end{equation}
Note that within this class of trial functions one always obtains
a finite value for $1/z^*$.

The amount of charge that can be adsorbed by the colloid is given by
$\alpha^* f N=(c_1/3 c_2) \; Q$, and is equal to $(16/15) \;Q \simeq 1.07 \;Q$ 
for the above
choice of trial function, which indeed suggests an overcharging, although
considerably smaller than reported in Ref. \cite{GS}. However, one can
see that this prediction is strongly dependent on the choice of the trial
wave function, and thus not robust. For example, a Gaussian wave function of the form
$\psi_z({\bf r})=(z^2/2 \pi)^{3/4} e^{-z^2 r^2/4}$ yields the above
results with $c_0=1/32 \sqrt{2}$, $c_1=2 \sqrt{2}/\sqrt{\pi}$, and 
$c_2=1/\sqrt{\pi}$. In this case, an adsorption of 
$(2 \sqrt{2}/3) \;Q \simeq 0.943 \;Q$ charges is 
predicted, which indicates an ``undercharging!'' 


Finally, I note that a finite size of the colloidal particle (a hard core),
and end effects due to finite length of the chain, have recently shown
to lead to overcharging \cite{Junior}. The overcharging is reduced as
the size of the particle decreases, and, interestingly, 
changes into a slight undercharging as it goes to zero \cite{Junior}.

In conclusion, I have shown that a variational approach can not be used
to unambiguously determine the degree of charging of a pointlike 
colloidal particle by an oppositely charged flexible polyelectrolyte.

It is a pleasure to acknowledge stimulating discussions 
with A. Yu. Grosberg, C. Jeppesen, E. Mateescu, and P. Pincus.
This research was supported in part by the National Science
Foundation under Grants No. PHY94-07194, and DMR-93-03667. 
\vskip0.45cm

\noindent Ramin Golestanian 

$\mbox{\rm Institute for Theoretical Physics}$

$\mbox{\rm  University of California}$

$\mbox{\rm  Santa Barbara, CA 93106-4030}$

\end{document}